\documentclass[english,prd,nofootinbib,preprintnumbers,twocolumn,showpacs]{revtex4}

\usepackage[utf8]{inputenc}
\usepackage[T1]{fontenc}
\usepackage{lmodern}  

\usepackage{amsmath,amssymb,amsfonts}
\usepackage{graphicx,epsfig,epstopdf}
\usepackage{color}
\usepackage[usenames,dvipsnames,svgnames]{xcolor}
\usepackage{adjustbox}
\usepackage{subfigure}
\usepackage{float}
\usepackage{appendix}
\usepackage{multirow}
\usepackage{cancel}
\usepackage{enumitem}
\usepackage{dcolumn}
\usepackage{bm}

\usepackage[colorlinks=true,
            linkcolor=red,
            urlcolor=gray,
            citecolor=blue]{hyperref}

\usepackage{graphics,psfrag,rotating}

\def\3nab{\tilde{\nabla}}

\def\be {\begin{equation}}
\def\ee {\end{equation}}
\def\ba {\begin{align}}
\def\ea {\end{align}}

\def\bc {\begin{center}}
\def\ec {\end{center}}
\def\case#1/#2{\frac{#1}{#2}}

\newcommand{\bver}{\begin{verbatim}}
\newcommand{\ever}{\end{verbatim}}

\newcommand{\bea}{\begin{eqnarray}}
\newcommand{\eea}{\end{eqnarray}}
\newcommand{\beaa}{\begin{eqnarray*}}
\newcommand{\eeaa}{\end{eqnarray*}}

\def\case#1/#2{\textstyle\frac{#1}{#2}}

\begin{document}

\title{Asymptotic Schwarzschild solutions in $f(R)$ gravity  \\ and their observable effects on the photon sphere of black holes}

\author{
Miguel Aparicio Resco 
}
\email{miguel.aparicio@upm.es}
\affiliation{Departamento de Matemática Aplicada a la Ingeniería Industrial, Escuela Técnica Superior de Ingeniería y Diseño Industrial, Universidad Politécnica de Madrid, Madrid, E-28012, Spain}




\begin{abstract} 

We investigate asymptotic Schwarzschild exterior solutions in the context of modified gravity theories, specifically within the framework of $f(R)$ gravity, where the asymptotic behavior recovers the standard Schwarzschild solution of General Relativity. Unlike previous studies that rely mainly on analytical approximations, our approach combines asymptotic analysis with numerical integration of the underlying differential equations. Using these solutions, we analyze strong lensing effects to obtain the photon sphere radius and the corresponding capture parameter. Considering rings produced by total reflection, we define the photon sphere width as the difference between the first total reflection and the capture parameter; and study how it is modified in the $f(R)$ scenario. Our results show that the photon sphere width increases in the presence of $f(R)$-type modifications, indicating deviations from GR that could be observable in the strong-field regime.

\end{abstract} 

\maketitle

\section{Introduction}

Modified gravity theories have emerged as natural extensions of General Relativity (GR) aiming to address cosmological and astrophysical phenomena such as the late-time accelerated expansion of the Universe, the nature of dark matter, and possible deviations from GR in the strong-field regime \cite{Clifton:2012, Sotiriou:2008, DeFelice:2010, Joyce:2016, Berti:2015}. Among these proposals, $f(R)$ gravity, where the Einstein-Hilbert Lagrangian is generalized to a function of the Ricci scalar $R$, is one of the simplest and most widely studied modifications. These theories have been extensively used to investigate compact objects and black hole solutions, providing a framework to explore how higher-order curvature corrections might modify the Schwarzschild (SW) and Kerr solutions of GR \cite{Clifton:2006ug, delaCruz-Dombriz:2009pzc, Nzioki:2009av, Astashenok:2017dpo, Campbell:2024ouk, AparicioResco:2016xcm}.\\

In the context of black holes, one of the most fundamental geometric features is the photon sphere, defined as the unstable circular null geodesic around the black hole \cite{Chael:2021rjo, Wielgus:2021peu, Sneppen:2021taq, Carballo-Rubio:2024uas}. In GR, the photon sphere radius is $3/2$ times the SW radius \cite{Chandrasekhar:1983, Wald:1984, Perlick:2004, Cardoso:2019}. The dynamics of null geodesics near the photon sphere are of particular relevance, as they determine observable phenomena such as black hole shadows \cite{EventHorizonTelescope:2019dse, EventHorizonTelescope:2019ggy, EventHorizonTelescope:2022wkp, Zhu2018, Lockhart:2022rui, Broderick:2022tfu, Galison:2024bop, Lupsasca:2024xhq}. The deflection of light near the photon sphere radius, and the transition between scattered and captured photons, are characterized by the so-called \emph{capture parameter}, which identifies the critical value of the impact parameter beyond which light is absorbed by the black hole. In GR it is known that this capture parameter is $3\sqrt{3}/2$ times the SW radius \cite{Chandrasekhar:1983, Wald:1984, Perlick:2004, Cardoso:2019}. As the impact parameter approaches the capture value, photons undergo an increasing number of strong deflections, and in the limiting case the trajectory asymptotically performs an infinite number of loops around the photon sphere. In particular, there are some deflection angles that are equally spaced on a logarithmic scale where light is totally reflected or transmitted \cite{Sneppen:2021taq}. Precise studies of these quantities are not only theoretically motivated but also essential for interpreting high-angular-resolution black-hole observations, such as those performed by the Event Horizon Telescope (EHT) \cite{EventHorizonTelescope:2019dse, EventHorizonTelescope:2022wkp}. In this regard, a methodological controversy arose in 2022 regarding the claimed detection of the photon sphere in M87* using the 2017 EHT data. Broderick et al.~\cite{Broderick:2022tfu} modeled a thin ring component consistent with the presence of a photon sphere ring, while Lockhart and Gralla~\cite{Lockhart:2022rui} argued that such evidence was not statistically robust and that the data could be explained equally well without invoking this feature. The consensus that has since emerged is that the EHT has not achieved a direct detection of the photon rings. Nevertheless, simulation studies predict that the next-generation EHT (ngEHT), operating at 230/345 GHz with enhanced baseline coverage and sensitivity, will reach the angular resolution required for a robust measurement \cite{Tiede:2022ngEHT}.\\

Recent literature has examined black hole optics in several modified gravity theories \cite{Koch:2025gaw, Diaz-Guerra:2025jip, Olmo:2025ctf}. In $f(R)$ gravity, Addazi, Capozziello and Odintsov found that instabilities in photon circular orbits lead to a double-exponential sensitivity of the black hole shadow; they showed how such chaotic solutions modify the photon sphere radius and the capture impact parameter \cite{Addazi2021}. Nojiri and Odintsov extended this by deriving the field equations for general static, spherically symmetric configurations in $f(R)$ gravity and used it to obtain the photon-sphere and shadow radii in Schwarzschild–de-Sitter backgrounds \cite{Nojiri2024}. Yue, Xu and Tang employed a Konoplya–Zhidenko deformation rule to model deviations from GR; they showed that increasing the deformation parameter $\varepsilon$ enlarges the photon sphere radius and critical impact parameter and used EHT data to constrain $\varepsilon$ \cite{Yue2025}. Focusing on strong gravitational lensing, Naskar, Molla and Debnath studied black holes in $f(R)$ Euler–Heisenberg Gravity’s Rainbow and reported that a larger Euler–Heisenberg parameter increases the photon-sphere radius and capture parameter whereas electric charge has the opposite effect \cite{Naskar2025}. Finally, Jafarzade, Bazyar and Jamil analyzed shadows and light deflection in $f(R)$–ModMax gravity; they found that matching the Event Horizon Telescope observations requires $f_{R_0}<-1$ for anti–de~Sitter black holes and $f_{R_0}>-1$ for de-Sitter black holes \cite{Jafarzade2024}.\\

In this work, we analyze black hole solutions in $f(R)$ gravity that smoothly connect to the Schwarzschild solution in the exterior region. In a recent work, Scali and Piattella \cite{ScaliPiattella2024} construct a class of asymptotically Schwarzschild solutions in $f(R)$ gravity by perturbatively solving the modified Einstein equations and recovering the corresponding $f(R)$ a posteriori, which turns out to be non-analytic at $R=0$. However, we will focus on a phenomenological model $f(R) = R + aR^2$ by assuming analytic $f(R)$ and that the Ricci scalar remains small in the relevant regime \cite{delaCruz-Dombriz:2009pzc, Nzioki:2009av}. We show that, in the limit of large radii, the Schwarzschild solution is asymptotically recovered. This model exhibits two distinct asymptotic behaviors: for $a>0$, the solutions decay exponentially, whereas for $a<0$, they show damped oscillations \cite{AparicioResco:2016xcm}. Starting from the field equations, we obtain the asymptotic behaviour of the metric functions analytically. These expressions are then used to initialise the numerical integration: at a sufficiently large radius, where the asymptotic regime is valid, we set the values of the metric functions and their derivatives according to the analytic expansion. In this way, the asymptotic analytic solutions provide the initial conditions for the fourth-order Runge-Kutta integration toward smaller radii. This guarantees that the resulting numerical solutions display the desired asymptotic properties.\\

Once the numerical solutions are obtained, we use them to compute the photon sphere radius and the capture parameter in these modified spacetimes. To this end, we generalize the standard GR calculation to a generic static, spherically symmetric metric and derive the corresponding generalized Binet equation. This equation is then solved numerically to determine the first value of the impact parameter leading to total reflection, i.e.\ when the deflection angle equals $\pi$, so that the photon returns along the incident direction. Together with the capture parameter, we define the effective width of the photon rings, which we shall briefly refer to as the \emph{photon sphere width}: the radial distance between the impact parameter of the first total reflection and that of infinitely many reflections, i. e. the capture parameter. This observable is particularly relevant, as it is a parameter that indicates the width of the photon ring structure that could be measured by future high-precision experiments, such as the Einstein Telescope and next-generation VLBI arrays \cite{Galison:2024bop, Lupsasca:2024xhq}.\\

Finally, the article is organized as follows. In Sec.~\ref{f(r) exter}, we briefly review the formulation of $f(R)$ gravity, present the system of equations for static, spherically symmetric metrics, study their asymptotic behavior, and solve the metric in the different cases. In Sec.~\ref{pho_sphe_capture}, we revisit the standard calculation of light deflection in black holes within GR and extend it to compute the photon sphere radius and the capture radius in the context of modified gravity. In Sec.~\ref{photon_width}, we investigate the strong-deflection limit in GR, briefly review the conditions for total reflection and transmission, and define the photon sphere width. In Sec.~\ref{pho_f_R}, we compute the photon sphere observables (radius, capture parameter, and width) using the $f(R)$ solutions obtained in Section II. Finally, Sec.~\ref{conclusions} summarizes our main results and discusses possible future directions.

\section{Perturbed Schwarzschild exterior solutions in $f(R)$}\label{f(r) exter}

The aim of this section is to present a numerical framework to obtain static, spherically symmetric solutions in $f(R)$ gravity. We also require that these solutions asymptotically recover the GR behavior. To this end, we first study the differential equation system to ensure the conditions required to recover the standard Schwarzschild solution at large radii. Then we obtain analytically the asymptotic solution to impose at large radii for the numerical resolution.\\

We start by considering the Einstein-Hilbert Lagrangian for a space-time with scalar curvature $R$ to a generic function $f(R)$, so the gravitational action becomes:
\begin{equation}
S = \frac{1}{2\kappa} \, \int d^4 x \, \sqrt{-g} \, \left( R + f(R)\right),
\end{equation}
where $\kappa = 8\pi G /c^4$. By varying with respect to the metric we obtain the field equations,
\begin{align}\label{fequations}
R_{\mu\nu} -& \frac{1}{2} R\, g_{\mu\nu} = \frac{1}{1 + f_R} \bigg[ -\kappa T_{\mu\nu} - \nabla_\mu \nabla_\nu f_R \notag \\
&\quad + g_{\mu\nu} \nabla^\alpha \nabla_\alpha f_R + \frac{1}{2} \left( f(R) - R f_R \right) g_{\mu\nu} \bigg],
\end{align}
where $f_R \equiv df(R)/dR$, and equivalent definition for higher derivatives $f_{RR}$, $f_{RRR}$, etc. The energy momentum tensor is defined as,
\begin{equation}
T_{\mu \nu} = \frac{2}{\sqrt{-g}} \, \frac{\delta \left( {\cal L}_{matter} \sqrt{-g} \right)}{\delta g^{\mu \nu}},
\end{equation}
for our case, we will consider vacuum solutions, so $T_{\mu \nu} = 0$. We consider the metric form for a static and spherically symmetric four-dimensional space-time,
\begin{equation}\label{metric}
ds^2 = B(r) \, dt^2 - A(r) \, dr^2 - r^2 \left( d\theta^2 +\sin^2 \theta \, d\phi^2 \right).
\end{equation}
Then using metric (\ref{metric}) on field equations (\ref{fequations}) we obtain the following system of differential equations for matter vacuum,
\begin{align}\label{system1}
\frac{dA}{dr} &= \frac{2rA}{3(1 + f_R)} \left[\frac{A}{2} R - f_R \left( \frac{A}{2} R + \frac{3}{2rB} \frac{dB}{dr} \right)  \right. \nonumber \\
&\quad \left. - \frac{3}{2rB} \frac{dB}{dr} + Af(R) - \left( \frac{3}{r} + \frac{3}{2B} \frac{dB}{dr} \right) f_{2R} \frac{dR}{dr} \right],
\end{align}
\begin{align}\label{system2}
\frac{d^2B}{dr^2} &= \frac{1}{2} \frac{dB}{dr} \left( \frac{1}{A} \frac{dA}{dr} + \frac{1}{B} \frac{dB}{dr} \right) + \frac{2B}{rA} \frac{dA}{dr} - \frac{2B}{(1 + f_R)} \nonumber \\
&\quad \left[ \frac{A}{2} R - \left( \frac{1}{2B} \frac{dB}{dr} + \frac{2}{r} \right) f_{2R} \frac{dR}{dr} + \frac{A}{2} f(R) \right],
\end{align}
\begin{align}\label{system3}
\frac{d^2R}{dr^2} &= \frac{dR}{dr} \left( \frac{1}{2A} \frac{dA}{dr} - \frac{1}{2B} \frac{dB}{dr} - \frac{2}{r} \right) - \frac{f_{3R}}{f_{2R}} \left( \frac{dR}{dr} \right)^2 \nonumber \\
&\quad + \frac{A}{3f_{2R}} \left[ (1 - f_R) R + 2f(R) \right].
\end{align}
We are interested in perturbed solutions around the exterior Schwarzschild (SW) solution of GR, where $R \sim 0$ compared to the characteristic curvature scales of the problem. In this situation we can expand a generic $f(R)$ theory at second order so $f(R) \approx R + a \, R^2$. As we want to recover SW, following the parametrization used in \cite{AparicioResco:2016xcm}, we rewrite $A(r)$ and $B(r)$ functions as,
\begin{equation}\label{Bfun}
B(r) \equiv 1 - \frac{r_S \, \left[ 1 + m(r) \right]}{r},
\end{equation}
\begin{equation}\label{Afun}
A(r) \equiv \frac{1+U(r)}{B(r)},
\end{equation}
where $r_S = 2GM/c^2$ in such a way $U(r)$ and $m(r)$ are dimensionless functions that reduces to zero for the SW asymptotic solution. Finally, since $r$, $R(r)$ and $a$ are dimensional quantities; we recast them as dimensionless parameters. We will express $r$ in units of $r_S$. We redefine $R(r)$ as $P(r) = 2\,a\,R(r)$ since $f_R (R) = 2\,a\,R(r)$ measures dimensionless deviations with respect to GR. Finally, the ratio $r/\sqrt{3a}$ naturally emerges in the asymptotic solutions, so we define $\alpha = \sqrt{3a}/r_S$. Then we introduce the following definitions:
\begin{equation}\label{def1}
r = r_S \,  x,
\end{equation}\\
\begin{equation}\label{def2}
R(r) = \frac{3 P(r)}{2 \alpha^2 r_S^2},
\end{equation}
\begin{equation}\label{def3}
a = \frac{\alpha^2 r_S^2}{3}.
\end{equation}
Using expressions (\ref{Bfun} - \ref{def3}) and considering $f(R) = R + a \, R^2$ on equations (\ref{system1}-\ref{system3}) we obtain,
\begin{widetext}
\begin{equation}\label{system_non_lin1}
U' = \frac{(1 + U)}{\alpha^2 (x -1 - m) (2 + P)} \, \left[ x^2 P (1 + U) + \alpha^2 \left(1 - 2x + m + x \, m' \right) P' \right],
\end{equation}
\begin{equation}\label{system_non_lin2}
m'' = \frac{
x^2 \, P \, (1 + U) \, \left[ 4x - 6 - 3(1 - x) \, P - 3 \, m \,(2 + P) + 2 \, x \, m' \right]
- 2 \, \alpha^2 \, \left[1 + m - x \, m'\right] \, \left[3 - 4x + 3 m + x \, m' \right] \, P'
}{
4 \, \alpha^2 \, x \, (x -1 - m) \, (2 + P)
},
\end{equation}
\begin{equation}\label{system_non_lin3}
P'' = \frac{
\left[4 + 2 \, P + x \, P'\right] \left[ x^2 \, P \,(1 + U) + \alpha^2 \, \left(1 - 2x + m + x \, m' \right) P' \right]
}{
2 \, \alpha^2 \, x \, (x -1 - m) \, (2 + P)
},
\end{equation}
\end{widetext}
where primes denote derivatives with respect to $x$. Now we consider a perturbation around SW so $U$, $m$, $P$ and their derivatives are much less than $1$ then we linearize the system above yielding,
\begin{equation}\label{lin1}
U' = \frac{1}{2\,(x-1)} \, \left[ \frac{x^2}{\alpha^2} \, P + \left( 1-2x \right) \, P' \right],
\end{equation}
\begin{equation}\label{lin2}
m'' = \frac{1}{4\,x\,(x-1)} \, \left[ \frac{x^2}{\alpha^2} \, \left( 2x-3 \right) \, P + \left( 4x-3 \right) \, P' \right],
\end{equation}
\begin{equation}\label{lin3}
P'' = \frac{1}{x\,(x-1)} \, \left[ \frac{x^2}{\alpha^2} \, P + \left( 1-2x \right) \, P' \right].
\end{equation}

As we can see in (\ref{lin1}-\ref{lin3}) and in the nonlinear system (\ref{system_non_lin1}-\ref{system_non_lin3}), there is a fixed point at $P = P' = m' = 0$. This point corresponds to the Schwarzschild solution when $m = U = 0$. On the other hand, the Schwarzschild geometry also requires $P = 0$, since $P$ parametrises deviations from the Ricci-flat condition $R=0$. This has been proved in other references \cite{delaCruz-Dombriz:2009pzc} but we want to analyze solutions that tend to SW at large radii. For doing that, we will solve equations (\ref{lin1}-\ref{lin3}) considering $x\gg1$ so we keep leading order terms for $P$ and $P'$ at that limit,
\begin{equation}\label{limit1}
U' \approx \frac{x}{2\,\alpha^2} \, P - P',
\end{equation}
\begin{equation}\label{limit2}
m'' \approx \frac{x}{2\,\alpha^2} \, P + \frac{1}{x} \, P',
\end{equation}
\begin{equation}\label{limit3}
P'' \approx \frac{1}{\alpha^2} \, P - \frac{2}{x} \, P'.
\end{equation}

From equation (\ref{limit3}) and accordingly with \cite{AparicioResco:2016xcm}, we can see two different behaviors depending on the sign of $a$. If $a>0$ then $\alpha^{2}>0$ and we obtain exponentially decaying solutions, while for $a<0$ one has $\alpha^{2}<0$ and the function $P(x)$ develops damped oscillations. It is worth recalling that a well-known viability condition for metric $f(R)$ gravity is the absence of the Dolgov-Kawasaki instability, which arises when curvature perturbations in a matter background acquire a negative effective mass squared. This phenomenon is controlled by the sign of $f_{RR}\equiv d^{2}f/dR^{2}$, and stability requires $f_{RR}>0$. For the quadratic approximation considered here, $f(R)=R+aR^{2}$, one has $f_{RR}=2a$, so the Dolgov-Kawasaki criterion would translate into $a>0$ if this truncated model were taken as fundamental. However, our analysis is perturbative around vacuum and does not aim to describe regimes of high curvature, where the full (non-perturbative) $f(R)$ Lagrangian may indeed satisfy $f_{RR}>0$ even if the effective low-curvature coefficient $a$ is negative. Therefore, the oscillatory branch ($a<0$) should not be regarded as necessarily incompatible with global stability, but rather as a possible exterior behavior emerging from an underlying stable $f(R)$ theory. The exponentially decaying branch ($a>0$), on the other hand, naturally aligns with the usual stability requirement.\\

We analyze both cases in the next subsections and, once we obtain the corresponding asymptotic analytic solutions, we use these expressions as initial conditions at a sufficiently large radius for the numerical integration of the system. In this way, we construct the solution that satisfies the required asymptotic behaviour.

\begin{figure*}[t]
    \centering
    \includegraphics[width=0.95\textwidth]{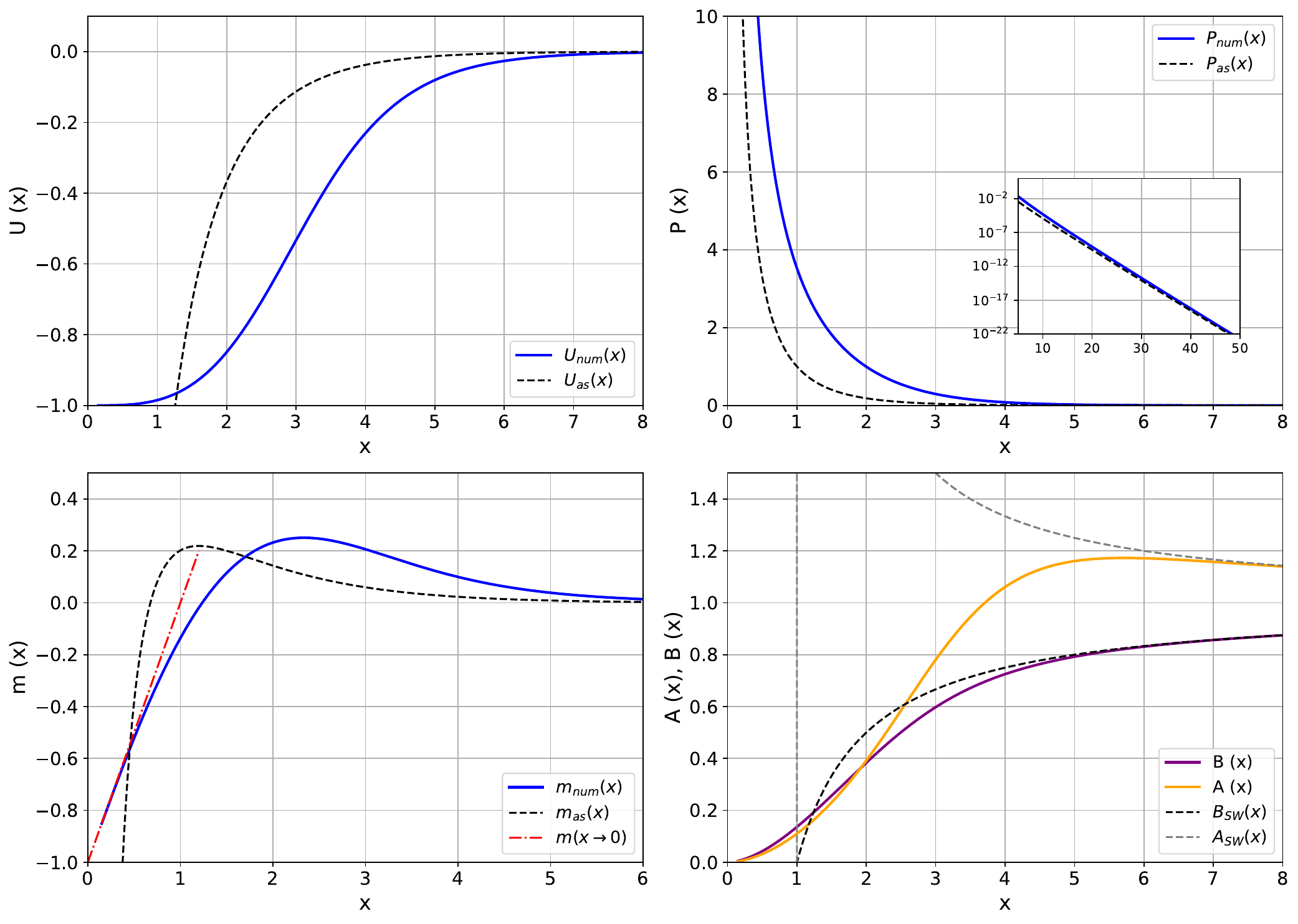}
    \caption{From left to right and upper to lower: numerical and asymptotic solutions for $U(x)$, $P(x)$, $m(x)$; and numerical solutions for metric functions $A(x)$ and $B(x)$ with the corresponding functions for the SW case $A_{SW}(x)$ and $B_{SW}(x)$. We can see that result are consistent with the asymptotic behavior. At lower radii both metric functions tend to zero and the function $P(x)$ diverges. In the left lower panel we also plot the behavior for low $x$ which is $m(x\to 0) = x - 1$. These results are obtained using $\alpha = 1$ and $P_{\alpha} = 1$.}
    \label{fig_result_a_Pa_1}
\end{figure*}

\subsection{Results for $a>0$}

As shown in system (\ref{limit1}-\ref{limit3}), we can solve (\ref{limit3}) and then use the result on equations (\ref{limit1}) and (\ref{limit2}). We need to impose the asymptotic conditions that are $U(x\to \infty) = 0$, $P(x\to \infty) = 0$, $P'(x\to \infty) = 0$, $m(x\to \infty) = 0$ and $m'(x\to \infty) = 0$. Note that if $m'(x\to \infty) = \mathrm{const.} \neq 0$ we obtain $B(x\to \infty) = \mathrm{const.} \neq 1$ that we can reabsorb in $t$ definition. Taking this into account, we consider only solutions of (\ref{limit3}) as decreasing exponential functions so,
\begin{equation}\label{P_x_exp}
P_{as}^{+}(x) = P_{\alpha} \, \frac{\alpha}{x} \, {\rm e}^{1-\frac{x}{\alpha}}, 
\end{equation}
where we have defined $P_{\alpha} = P(x=\alpha)$ which is, in addition with $\alpha$, a parameter for the solution that describes the perturbation of Ricci scalar from zero. This parameter can be seen as an effective dark energy fluid related with the modified gravity model. Using the result from (\ref{P_x_exp}), we can integrate (\ref{limit1}) and (\ref{limit2}) to obtain,
\begin{equation}
U_{as}^{+}(x) = -\frac{P_{\alpha}}{2} \, \left( 1 + \frac{2\, \alpha}{x} \right) \, {\rm e}^{1-\frac{x}{\alpha}},
\end{equation}
\begin{align}\label{mass_a_pos_exp}
m_{as}^{+}(x) = \frac{P_\alpha}{2} \left[ \left( 1 + \alpha - \frac{\alpha}{x} \right) \, {\rm e}^{1-x/\alpha} + \frac{x}{\alpha} \, {\rm e} \, \text{Ei}\left( -\frac{x}{\alpha} \right) \right],
\end{align}
where $\mathrm{Ei}\left( x \right)$ is the exponential integral. Once we have the asymptotic solutions (\ref{P_x_exp}-\ref{mass_a_pos_exp}), we solve numerically the system (\ref{system1}-\ref{system3}) using a fourth-order Runge-Kutta algorithm. For doing that we need [$m_{as \, 0}$, $m'_{as \, 0}$, $P_{as \, 0}$, $P'_{as \, 0}$, $U_{as \, 0}$] evaluated at some radius $x_0$, as $x = r/r_S$, a value of $x>>1$ imply a large value. We have also the parameter $\alpha$ which is a radii scale related to the SW deviation, for our analysis $\alpha < 2$. Based on this, we have shown that a value of $x_0 = 150$ is sufficient to evaluate the asymptotic solution without introducing significant error. In Figure \ref{fig_result_a_Pa_1} we plot the results for $U(x)$, $P(x)$, $m(x)$ and metric functions $A(x)$ and $B(x)$. We consider for this plot $\alpha = P_{\alpha} = 1$. As we can see, the value of $P(x)$ diverges for $x \to 0$ and both metric functions tend to zero which implies that $U(x \to 0) = -1$ and $m(x \to 0) = x-1$. Due to the definition of $m(x)$, we can read it as a correction to the mass of the black hole at large radii. This correction grows up to order $20\%$ at $x \approx 2$ and then decays.\\

\begin{figure}[H]
    \centering
    \includegraphics[width=\columnwidth]{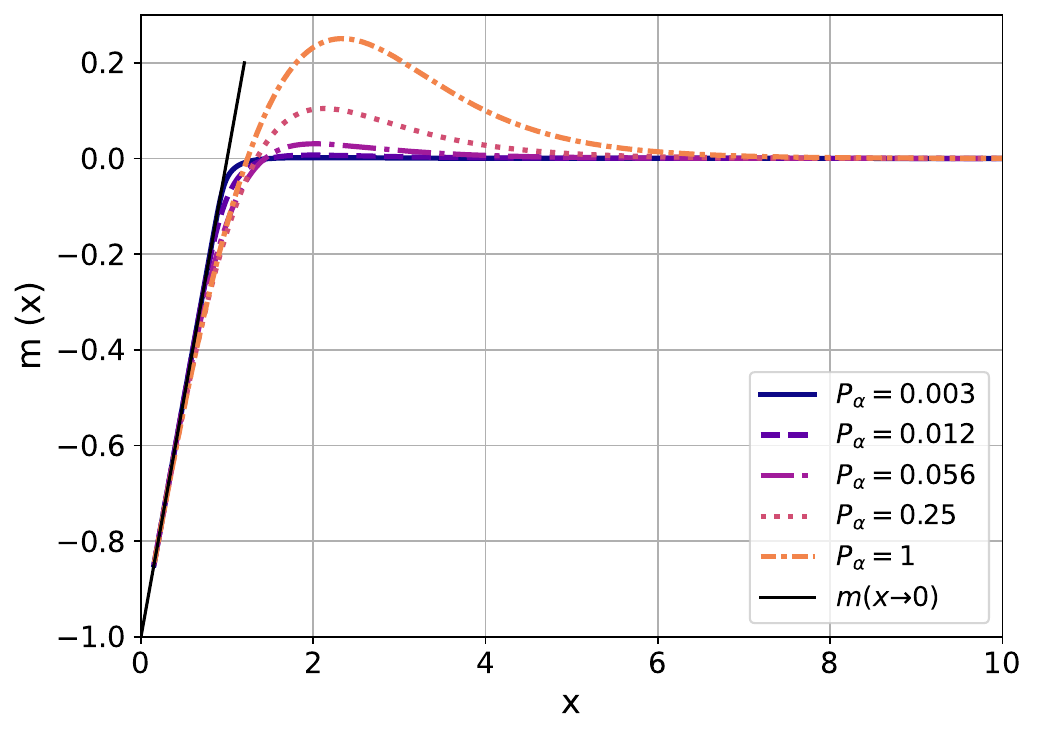}
    \caption{$m(x)$ for different values of $P_{\alpha}$ and with $\alpha = 1$ fixed. As we can see, all curves tends to $x-1$ when $x \to 0$. Increasing $P_{\alpha}$ makes the maximum of $m(x)$ bigger at the same radii.}
    \label{fig_result_a_fix}
\end{figure}

We explore now the dependence of the solutions with respect to $\alpha$ and $P_{\alpha}$. In Figure \ref{fig_result_a_fix} we fix the value of $\alpha = 1$ and move $P_{\alpha}$. As we can see, in the limit of $P_{\alpha} \to 0$ the function $m(x)$ tends to a zero constant up to $x=1$ in which $m(x) \to x-1$. This is due to the factor $(x-1-m)$ in the denominator of equations (\ref{system_non_lin1}-\ref{system_non_lin3}). Moreover, in the exact case of $P_{\alpha} = 0$ we recover SW solution and the system (\ref{system_non_lin1}-\ref{system_non_lin3}) is not able to integrate for values less than $x=1$. This suggests that the system is not appropriate to analyze solutions for $x \to 0$. Nevertheless, in this work we are interested in the photon sphere, so we focus on metric functions in the range $x>1$.\\

It is important to emphasize that the numerical solutions obtained here should only be regarded as valid in the regime where the Ricci scalar remains small, that is, for radii not too close to the region where the curvature begins to grow appreciably. As the curvature grows, the approximation $f(R) \simeq R + aR^2$ ceases to be reliable, and higher-order corrections in $R$ (such as $R^3$, $R^4$, etc.) would need to be included. Therefore, the divergent behavior of $R(r)$ and the metric functions at $x \ll 1$ is not to be interpreted as a physical prediction of the model but rather as an indication of the breakdown of the truncated expansion.

\begin{figure}[H]
    \centering
    \includegraphics[width=\columnwidth]{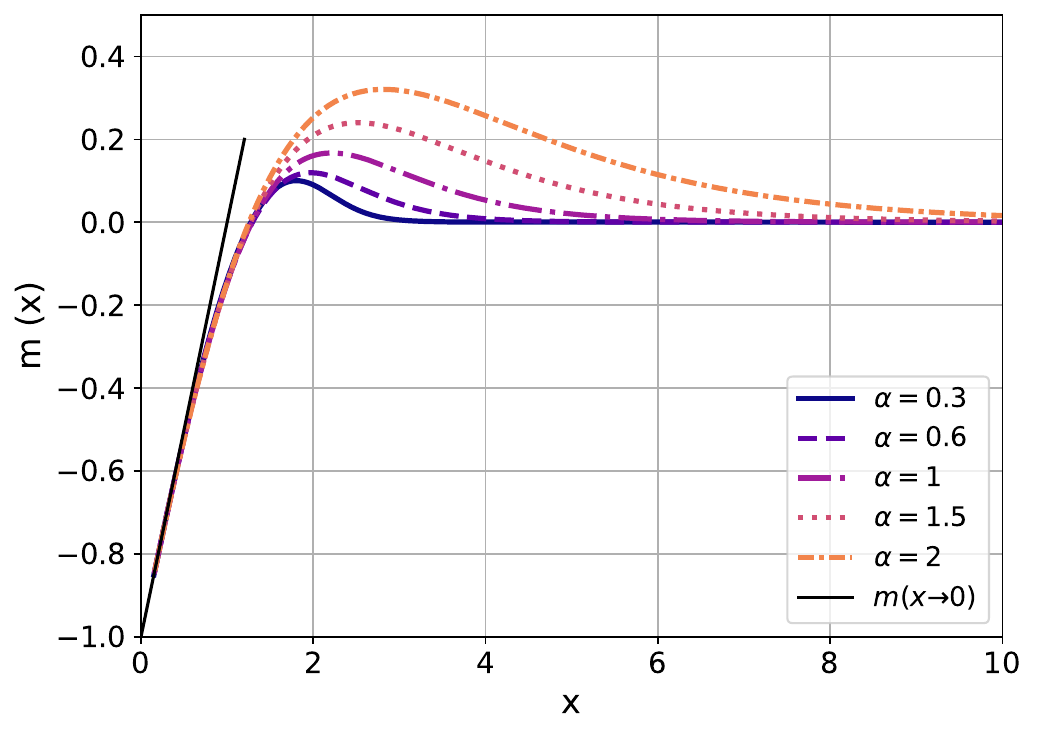}
    \caption{$m(x)$ for different values of $\alpha$ and with $P_{\alpha} = 0.5$ fixed. As we can see, all curves tends to $x-1$ when $x \to 0$. Increasing $\alpha$ makes the maximum of $m(x)$ bigger and moves it to larger radius values.}    
    \label{fig_result_Pa_fix}
\end{figure}

In Figure \ref{fig_result_Pa_fix} we fix the value of $P_{\alpha} = 0.5$ and move $\alpha$. Increasing $\alpha$ makes the correction to mass bigger and moves the maximum value for larger values of radii. This can be seen in the asymptotic solution (\ref{mass_a_pos_exp}) due to the exponential term ${\rm e}^{1-x/\alpha}$.\\

Finally, we have also explored $P_{\alpha}<0$, in this case there is a divergence for $P(x)$, $U(x)$ and $m(x)$ at some radii $x>1$ that tends to $1$ in the limit of $P_{\alpha} \to 0$. This effect is due to the term $(2+P)$ on denominators of (\ref{system_non_lin1}-\ref{system_non_lin3}), as soon as $P(x) \to -2$ the solution diverges. Because these divergences prevent a physically meaningful interpretation of the resulting profiles, we have not analysed this case further.

\subsection{Results for $a<0$}

In this case the asymptotic systems can be rewritten as,
\begin{equation}\label{a_neg_sys1}
U' \approx -\frac{x}{2\,|\alpha|^2} \, P - P',
\end{equation}
\begin{equation}\label{a_neg_sys2}
m'' \approx -\frac{x}{2\,|\alpha|^2} \, P + \frac{1}{x} \, P',
\end{equation}
\begin{equation}\label{a_neg_sys3}
P'' \approx -\frac{1}{|\alpha|^2} \, P - \frac{2}{x} \, P',
\end{equation}
where we have considered explicitly that $\alpha^2 <0$ due to the definition of $\alpha$, so in the following we will work with $|\alpha|$ by considering explicitly the minus sign in the equations. Now, as can be seen in equation (\ref{a_neg_sys3}), $P(x)$ possesses damped oscillatory solutions.  In this situation the asymptotic conditions: $U(x\to \infty) = 0$, $P(x\to \infty) = 0$, $P'(x\to \infty) = 0$, $m(x\to \infty) = 0$ and $m'(x\to \infty) = 0$; can only be satisfied on average over length scales $\Delta x \gg |\alpha|$, but not exactly.\\

If we integrate equation (\ref{a_neg_sys3}) we have solutions of type $\sin(x/|\alpha|)/x$ and $\cos(x/|\alpha|)/x$, in principle both are possible. However, when we integrate equation (\ref{a_neg_sys2}) we see that $\sin(x)$ case gives a term $x\, \mathrm{Si}(x)$ which diverges at $x \to \infty$. So this means that the only possibility for $P(x)$ is of the form of $\cos(x)$ that results on $x\, \mathrm{Ci}(x)$ whose average is zero at large $x$:
\begin{equation}\label{P_x_osc}
P_{as}^{-}(x) = \frac{|\alpha| \, P_{\alpha}}{x} \, \cos \left( \frac{x}{|\alpha|} \right), 
\end{equation}
\begin{equation}
U_{as}^{-}(x) = -\frac{|\alpha| P_{\alpha}}{x} \, \cos\left(\frac{x}{|\alpha|}\right) - \frac{P_{\alpha}}{2} \, \sin\left(\frac{x}{|\alpha|}\right)
\end{equation}
\begin{align}\label{mass_a_pos_osc}
m_{as}^{-}(x) = &\, \frac{P_{\alpha}}{2} \, \left[ |\alpha| \cos\left(\frac{x}{|\alpha|}\right) + \sin\left(\frac{x}{|\alpha|}\right) \right. \nonumber \\
&\left.- \frac{|\alpha|}{x} \, \cos\left(\frac{x}{|\alpha|}\right) \right] - \frac{P_{\alpha}}{2} \frac{x}{|\alpha|}\, \mathrm{Ci}\left(\frac{x}{|\alpha|}\right).
\end{align}
Once we have the asymptotic solutions, we solve numerically the system using a fourth-order Runge-Kutta algorithm. As in the previous section, we use the asymptotic solutions to give initial conditions at $x_0 = 150$ and make the integration to lower radii.\\

\begin{figure*}[t]
    \centering
    \includegraphics[width=0.95\textwidth]{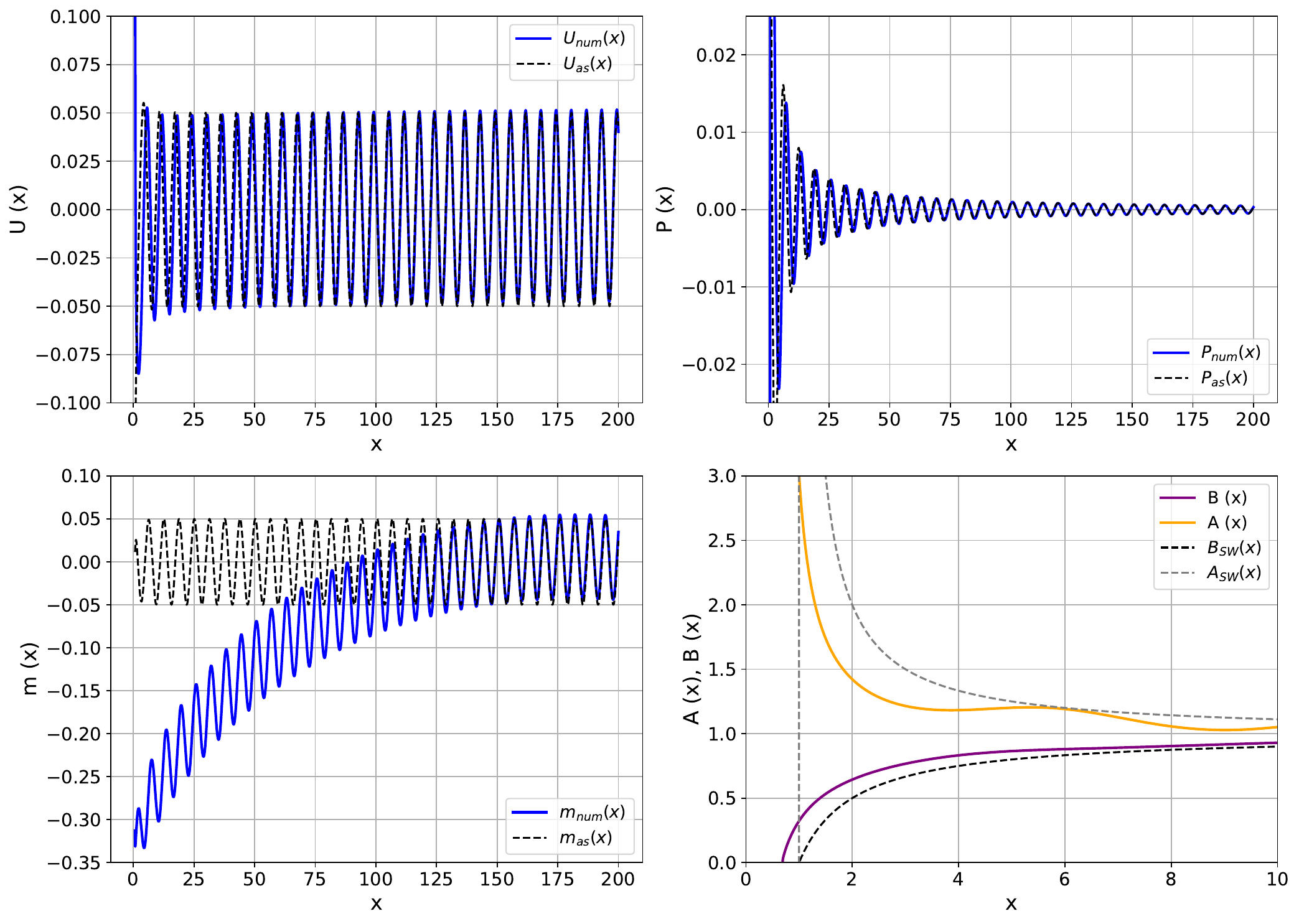}
    \caption{From left to right and upper to lower: numerical and asymptotic solutions for $U(x)$, $P(x)$, $m(x)$; and numerical solutions for metric functions $A(x)$ and $B(x)$ with the corresponding functions for the SW case $A_{SW}(x)$ and $B_{SW}(x)$. We can see that the numerical results are consistent with the asymptotic behavior. The metric functions at low radii are similar to SW with some oscillating pattern and with an event horizon lower than $x=1$. These results are obtained using $|\alpha| = 1$ and $P_{\alpha} = 0.1$.}
    \label{fig_result_a_Pa_osc}
\end{figure*}

We plot in Figure \ref{fig_result_a_Pa_osc} results for $|\alpha| = 1$ and $P_{\alpha} = 0.1$. These solutions resemble the oscillatory behavior reported in \cite{AparicioResco:2016xcm}. In this case, we can integrate from $x \gg 1$ down to the event-horizon region, where $A(x)$ diverges. In our solutions this divergence occurs simultaneously with $B(x) \to 0$, so both conditions identify the same radius.\\

An interesting feature emerges: these $f(R)$ black holes have a smaller horizon than the asymptotic SW mass would suggest. A distant observer infers the mass parameter $M$ from the asymptotic $1/r$ behaviour of the metric, with the associated Schwarzschild radius $r_{S}=2GM/c^2$ entering our formulation through the definition of the dimensionless coordinate $x = r/r_{S}$. This rescaling allows the numerical solutions to be expressed independently of the specific value of $M$, which simply sets the overall length scale. In General Relativity, the event horizon would lie at $r=r_{S}$, i.e.\ at $x=1$. However, in the $f(R)$ model considered here the modified field equations shift the location of the horizon to values $x<1$, even though the asymptotic observer still attributes the mass $M$ to the solution. Consequently, the photon-sphere radius is smaller and its width is modified.\\

\begin{figure}[H]
    \centering
    \includegraphics[width=\columnwidth]{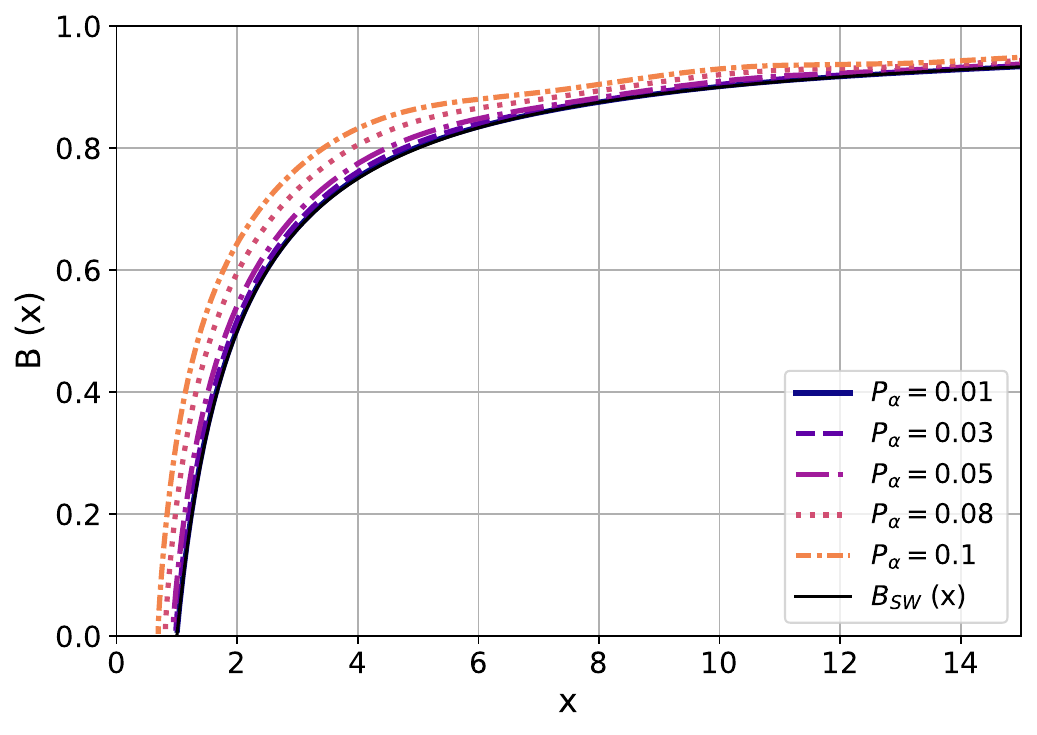}
    \caption{$B(x)$ for different values of $P_{\alpha}$ and with $|\alpha| = 1$ fixed. As we can see, bigger values of $P_{\alpha}$ make $B(x)$ greater than the SW solution at the same radii.}
    \label{fig_result_a_fix_osc}
\end{figure}

We now explore the dependence of the solution on $|\alpha|$ and $P_{\alpha}$. If we fix $P_{\alpha}$ and move $|\alpha|$ we find only modest differences for a range of $|\alpha| = 0.05 - 1$. For lower values of $|\alpha|$ the oscillations have shorter period and amplitude but they do not significantly alter the values of the metric function $B(x)$ with respect to the SW case. On the other hand, we can fix $|\alpha|$ and vary $P_{\alpha}$, results can be seen in Figure \ref{fig_result_a_fix_osc}. In the limit $P_{\alpha} \to 0$ we recover SW solution but at bigger values of $P_{\alpha}$ we get lower radius for the event horizon.

\begin{figure}[H]
    \centering
    \includegraphics[width=\columnwidth]{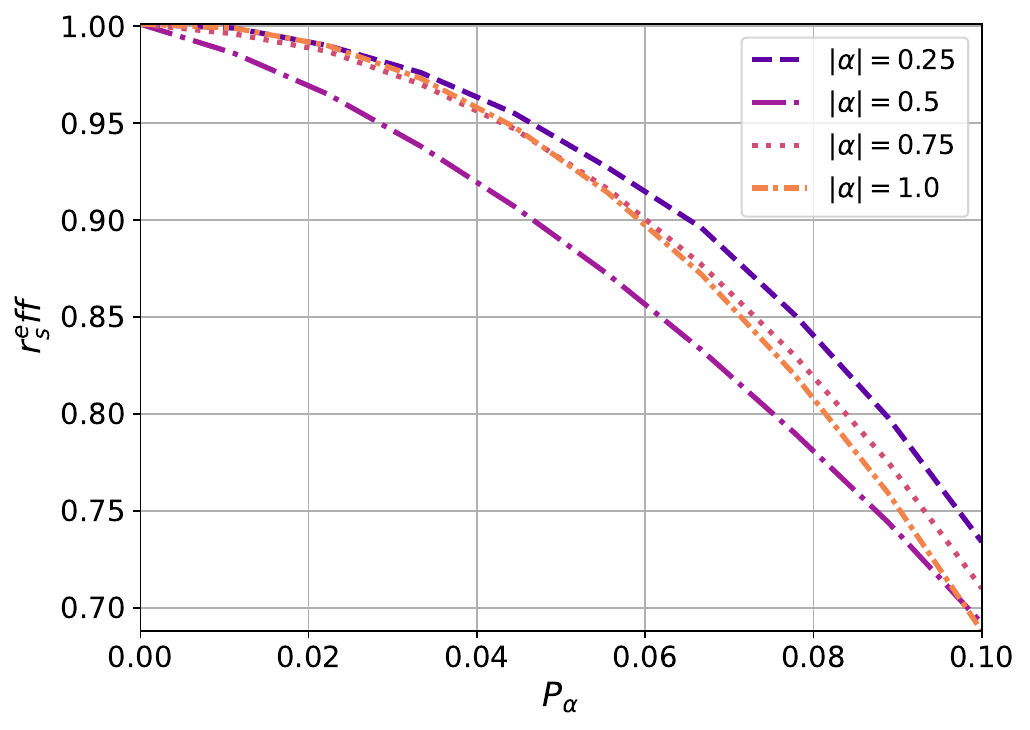}
    \caption{Values of the effective event horizon $r_S^{eff}$ in units of the asymptotic SW radius $r_S$ as a function of $P_{\alpha}$ and for different values of $|\alpha|$.}
    \label{fig_event}
\end{figure}

We can plot the effective event horizon for different values of $P_{\alpha}$ and $|\alpha|$, the results are seen in Figure \ref{fig_event}. As we can see, increasing $P_{\alpha}$ up to a value of $0.1$ makes the event horizon $30 \, \%$ lower. Different values of $|\alpha|$ give similar results from $|\alpha| = 0.25$ up to $1$.

As shown in Figure~\ref{fig_result_a_Pa_osc}, when $B(x) \to 0$ we find that $A(x) \to \infty$. From the metric in equation (\ref{metric}), it follows that the Ricci scalar remains finite at the effective event horizon. However, as $x \to 0$, the Ricci scalar grows rapidly and, as in the previous case, the approximation $f(R) \simeq R + aR^2$ is no longer valid. In this strong-curvature regime, higher-order corrections in $R$ would need to be taken into account to obtain physically meaningful results.

\section{Photon sphere radius and capture parameter} \label{pho_sphe_capture}

It is well known that a strong gravitational field bends light trajectories \cite{Sneppen:2021taq, Carballo-Rubio:2024uas}, and it is also well known that for static and spherically symmetric black holes there is a radius known as the photon sphere radius, $r_P$, at which light follow an unstable circular orbit. The impact parameter of a light ray that is captured at the photon sphere is referred to as the capture parameter $b_c$; and it can be obtained analytically in GR.
\\
\\
The aim of this section is to extend the derivation for an arbitrary static and spherically symmetric space-time given by (\ref{metric}). We start by considering the Lagrangian for a light trajectory in this space-time,
\begin{equation}\label{lagran}
\mathcal{L} = g_{\mu \nu} \, \dot{x}^{\mu} \, \dot{x}^{\nu} = B(r)\, \dot{t}^2 - A(r)\, \dot{r}^2 - r^2 \, \dot{\phi}^2 = 0,
\end{equation}
where dots denote derivatives with respect to the affine parameter of the trajectory, and we have considered that $\theta = \pi /2$ without loss of generality due to the spherical symmetry of the metric. We also have that geodesics for $\phi$ and $t$ can be written in terms of constants of motion,
\begin{equation}\label{const}
\dot{\phi} = \frac{L}{r^2}, \,\,\,\,\,\,\,\,\, \dot{t} = \frac{k}{B(r)},
\end{equation}
where $L$ and $k$ are constants related to the angular momentum and energy of the light respectively. Using expressions (\ref{const}) in Lagrangian (\ref{lagran}) we get a conservation equation,
\begin{equation}\label{cons}
k^2 = A(r) \, B(r) \, \dot{r}^2 + \frac{L^2}{r^2} \, B(r).
\end{equation}
We consider a light ray arriving from $r \to \infty$ with impact parameter $b$. Considering asymptotic behavior for metric functions $B(r \to \infty) = A(r \to \infty) = 1$ and considering that light is asymptotically free at $r \to \infty$ so $\dot{r}^2 (r \to \infty) = 1$, we can derive that $k=1$ from equation (\ref{cons}). 
\\
\\
To obtain the angular momentum $L$ of the ray, we consider that at $r \to \infty$ the trajectory is a line with impact parameter $b$ so in polar coordinates,
\begin{equation}
r_{\infty}(\phi) = \frac{b}{\sin \phi},
\end{equation}
then we consider equation (\ref{const}) for angular momentum $L$ and change $\dot{\phi} = (dr/d\phi)^{-1} \, \dot{r}$ so,
\begin{equation}
L = r^2 \, \left( \frac{dr}{d\phi} \right)^{-1} \, \dot{r} = - \frac{b \ \dot{r}}{\cos \phi},
\end{equation}
considering that at $r \to \infty$ we have $\phi \to 0$ (we consider without loss of generality that, in the $[x,y]$ plane, an incoming light ray cames from $r \to \infty$ at $[x \to \infty, b]$), and $\dot{r} \to -1$ because we consider an incoming path from $r \to \infty$ to zero. So we obtain that $L = b$ \cite{Chandrasekhar:1983, Wald:1984}. Then the conservation equation for the trajectory becomes,
\begin{equation}\label{cons2}
A(r) \, B(r) \, \dot{r}^2 + \frac{b^2}{r^2} \, B(r) = 1.
\end{equation}
To obtain the photon sphere radius $r_P$ we need the radial geodesic that can be obtained by differentiating (\ref{cons2}) with respect to the affine parameter. Then we impose the radii $r = r_P$ in which $\ddot{r} = \dot{r} = 0$ and we obtain,
\begin{equation}\label{photon_rad}
r_P = \frac{2\,B(r_P)}{B'(r_P)}, 
\end{equation}
here $r_P$ is expressed in units of the asymptotic SW radius $r_S$. As we can see, we obtain an implicit relation to obtain $r_P$. In the standard SW case $B(x) = 1-x^{-1}$ and we can check that relation (\ref{photon_rad}) gives $r_P =3/2$ which is the known value for GR.
\\
\\
Once we have this radius, we can obtain the capture parameter $b_c$ by considering a trajectory with this impact parameter that at $r \to r_P$ we obtain $\dot{r} \to 0$ from (\ref{cons2}) we obtain,
\begin{equation}\label{capture}
b_c = \frac{r_P}{\sqrt{B(r_P)}},
\end{equation}
considering the standard SW metric and $r_P = 3/2$ we recover the known value for capture parameter in GR $b_c = \sqrt{27}/2$ \cite{Chandrasekhar:1983, Cardoso:2019}.
\\
\\
Finally, we need the generalized Binet equation to solve numerically the trajectory. For doing that we will change variable $r$ to $u = r_S / r = 1/x$ in equation (\ref{cons2}),
\begin{equation}\label{cons_4}
A(u) \, B(u) \, \left( \frac{du}{d\phi} \right)^2 + u^2 B(u) = b^{-2},
\end{equation}
where we have used that $\dot{r}^2 = L^2 (du/d\phi)^2 = b^2 (du/d\phi)^2$, and $b$ have units of $r_S$. If we derive (\ref{cons_4}) with respect to $\phi$ and simplify, we obtain the generalized Binet equation,
\begin{equation}\label{binet_gen}
\frac{d^2 u}{d \phi^2} + \frac{1}{2} \left( \frac{1}{A} \frac{dA}{du} + \frac{1}{B} \frac{dB}{du} \right) \left( \frac{du}{d\phi} \right)^2 + \frac{u^2}{2AB} \frac{dB}{du} + \frac{u}{A} = 0,
\end{equation}
if we consider the standard SW metric we recover the relativistic Binet equation,
\begin{equation}\label{binet_GR}
\frac{d^2 u}{d \phi^2} + u = \frac{3}{2} \, u^2.
\end{equation}
To solve it numerically we impose that the ray comes from $x \to \infty$ that corresponds to $u \to 0$ at $\phi = 0$. We also need the initial condition for $du/d\phi$, this can be obtained with equation (\ref{cons_4}) by imposing that at $u \to 0$ we recover the SW metric, i.e. $A(u) \, B(u) = 1$, and then $du/d\phi = + b^{-1}$, with the $+$ sign implies that we consider the incoming ray.

\section{Photon sphere width}\label{photon_width}

In the previous section, we obtained the radius of the photon sphere and the capture parameter. As mentioned, the circular orbit is unstable since it corresponds to a maximum of the effective radial potential for null geodesics. This implies that light trajectories approaching the orbit will either be deflected away or fall into the black hole.
\\
\\
The observable effect will be a set of rings around the black hole, formed by light from surrounding sources deflected by its gravitational field. This effect has been studied before \cite{Galison:2024bop, Lupsasca:2024xhq, Chael:2021rjo, Wielgus:2021peu, Sneppen:2021taq}. In general, the pattern of deflected rings depends on the light sources surrounding the black hole, and sophisticated simulations are required to fit the experimental data.
\\
\\
In this work, we will focus on a particularly simple case: we will assume that the light sources lie along the line of sight of the observer. Therefore, we will concentrate on light trajectories that undergo total deflection and total transmission. This approach allows us to provide a simple definition of the observable width of the photon sphere.
\\
\\
In this section we will review the standard result for GR and define the photon sphere width $\delta_{P}$ for the SW solution. As can be seen in \cite{Sneppen:2021taq}, total reflection and transmission occur for impact parameters that exponentially approach the critical impact parameter, which corresponds to the capture parameter.

\begin{figure}[H]
    \centering
    \includegraphics[width=\columnwidth]{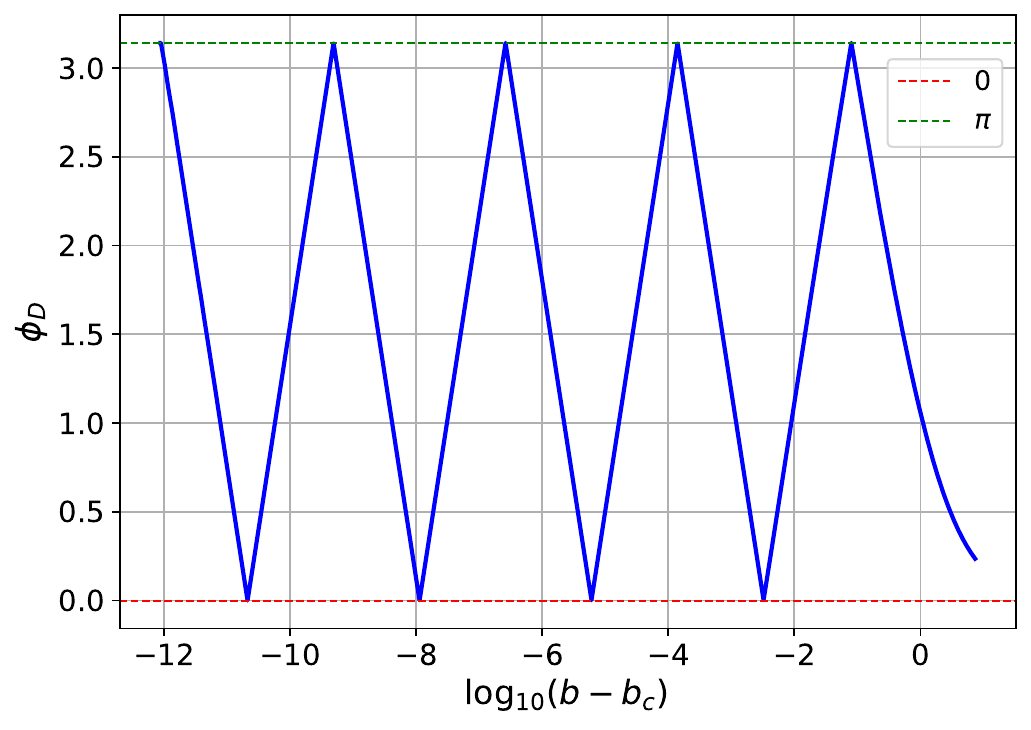}
    \caption{Values of the deflection angle $\phi_D$ as a function of the logarithmic difference of the impact parameter $b$ and the capture parameter $b_c$: $\log_{10 \left( b - b_c \right)}$. Total reflection occurs when $\phi_D = \pi$ and total transmission occurs when $\phi_D = 0$.}
    \label{fig_deflex_SW}
\end{figure}

We integrate equation (\ref{binet_GR}) with initial conditions $u(0)=0$ and $du/d\phi (0) = 1/b$ for a given impact parameter $b$. For each $b$, we integrate from $\phi = 0$ until one of these conditions occurs: $u(\phi_f) = 0$ or $u(\phi_f) \geq 1$. The first condition states that the light has been deflected at an angle $\phi_f$, while the second indicates that the light has fallen into the black hole.\\
\\
As we know that the light is always deflected for impact parameters $b>b_c$, we analyze those cases and calculate the deflection angle,
\begin{equation}
\phi_D = \left|\phi_f-2\pi \left\lfloor \frac{\phi_f}{2\pi} \right\rfloor - \pi \right|,
\end{equation}
where $\left\lfloor x \right\rfloor$ is the floor function. We plot in Figure \ref{fig_deflex_SW} the deflected angle as a function of the logarithmic difference of the impact parameter $b$ and the capture parameter $b_c$. When $\phi_D = \pi$ we have that light makes $N$ loops around the black hole and is then totally reflected by an angle $\pi$, for $N = 0, 1, 2, 3, 4$, from right to left in the figure. In addition, when $\phi_D = 0$, the light performs $N$ loops and is then fully transmitted for $N = 1, 2, 3, 4$.\\
\\
From the point of view of an observer far from the black hole, and considering only the incoming light paths aligned with the line of sight, the observable effect will be a set of rings caused by total reflection and transmission, with radii equal to the impact parameter in each case.
\\
\\
This set of rings is bounded by two radii: the capture impact parameter $b_c,$ which corresponds to the limit case where the light performs $N \to \infty$ loops and is then either totally reflected or transmitted, and $b_0,$ which is the radius associated with the first total reflection of light ($\phi_D = \pi$ for $N=0$). We therefore define the photon sphere width $\delta_P$ as,
\begin{equation}
\delta_P = b_0 - b_c.
\end{equation}
Considering the standard SW solution we obtain that $b_c = \sqrt{27}/2$ and $b_0 = 2.67848$ so the photon sphere width is $\delta_P^{SW} = 0.08040$ in units of the SW radius of the black hole. We can also estimate the angular size $\eta$ corresponding to this photon sphere width for a black hole of mass $M$ situated at a distance $D$ from the observer,
\begin{equation}
\eta \approx 6 \, \delta_P \, \frac{M}{D} \, 10^{11} \, \, \left( \mathrm{\mu as}  \right),
\end{equation}
where $M$ is given in units of solar masses, and $D$ in kilometers. For example, in the case of Sagittarius A* \cite{EventHorizonTelescope:2022wkp, Zhu2018} this angle is $\eta \approx 0.8 \, \mathrm{\mu as}$ and $\eta \approx 0.6 \, \mathrm{\mu as}$ for the case of Messier 87 (M87*) \cite{EventHorizonTelescope:2019dse, EventHorizonTelescope:2019ggy}. The resolution of the EHT is of the order of $20-25 \, \mathrm{\mu as}$, thus an improved resolution is required to directly observe this effect.

\section{Photon sphere observables\\ in $f(R)$ gravity} \label{pho_f_R}

Now we want to calculate the observables defined in the previous sections: the photon sphere radius $r_P$, the capture parameter $b_c$ and the photon sphere width $\delta_P$ in the context of $f(R)$ gravity. We will consider results from Section \ref{f(r) exter} for black-hole solutions in $f(R)$ gravity so we will consider the two cases: $a>0$ and $a<0$.\\

To measure the deviation of $\delta_P$ with respect to the standard value in GR, we will define,
\begin{equation}
\epsilon_{\delta_P} = 100 \cdot \frac{\delta_P - \delta_P^{SW}}{\delta_P^{SW}},
\end{equation}
which quantifies, in percent, the increase or decrease of the photon sphere width $\delta_P$ with respect to the GR value.

\subsection{Observables for $a>0$}

\begin{figure}[H]
    \centering
    \includegraphics[width=\columnwidth]{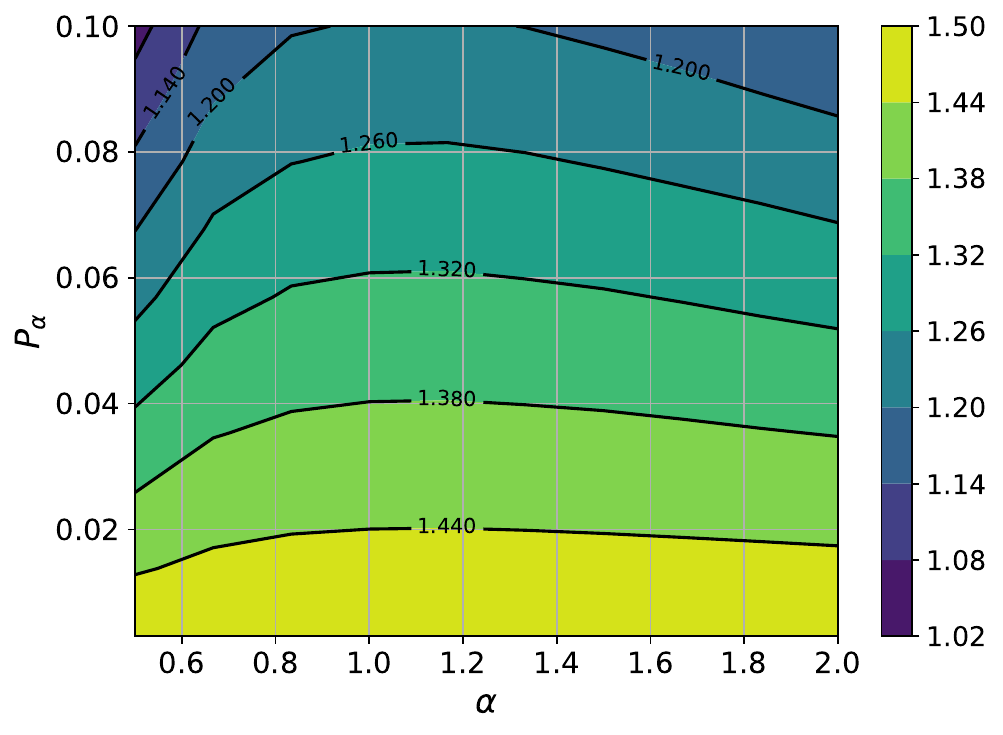}
    \caption{Values of the photon sphere radius $r_P$ for the case $a>0$ as a function of $\alpha$ and $P_{\alpha}$. As can be seen, for low values of the $f(R)$ correction we recover the standard radius $r_P = 3/2$ but as we increase $P_{\alpha}$ the photon sphere radius decreases.}
   
    \label{rp_exp_sol}
\end{figure}

In Figure \ref{rp_exp_sol} we plot photon sphere radius $r_P$ as a function of $\alpha$ and $P_{\alpha}$. We can see that $r_P$ approaches to $3/2$ for low values of $P_{\alpha}$. This is due to the fact that GR solution is recovered for $P_{\alpha} = 0$ because this parameter describes a dimensionless correction from GR (\ref{def2}). The dependence on $\alpha$ is milder, as it primarily sets the radial scale (in units of the SW radius) over which the corrections dominate. For a $f(R)$ correction up to $10 \%$, the photon sphere radius decreases down to $\sim1$.\\

\begin{figure}[H]
    \centering
    \includegraphics[width=\columnwidth]{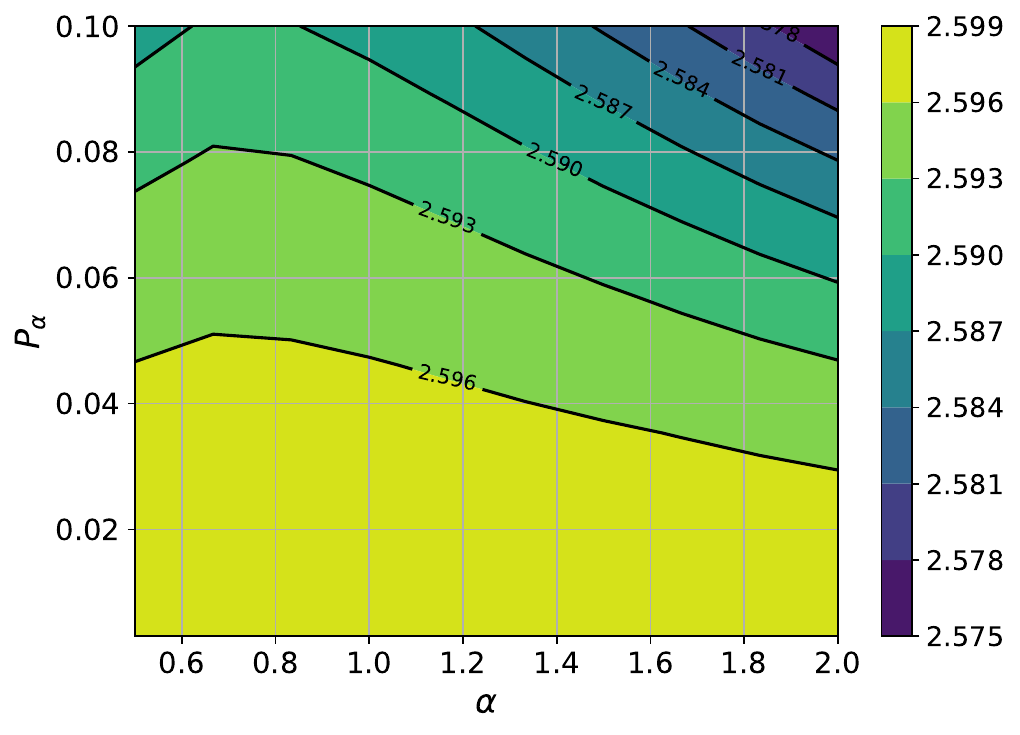}
    \caption{Values of the capture parameter $b_c$ for the case $a>0$ as a function of $\alpha$ and $P_{\alpha}$. As can be seen, as we increase $P_{\alpha}$ the capture parameter decrease with respect to standard value $b_c = \sqrt{27}/2$.}
    \label{bc_exp_sol}
\end{figure}

In Figure \ref{bc_exp_sol} we plot capture parameter $b_c$ as a function of $\alpha$ and $P_{\alpha}$. Similarly, the standard GR result is recovered when $P_{\alpha} \to 0$ and also this capture parameter decrease with respect to GR. However, in opposite of last result, the decrease of $b_c$ is not so pronounced as in the $r_P$ case. For a $f(R)$ correction up to $10 \%$, the capture parameter decrease from $2.598$ to $2.575$. This will have an effect on the photon sphere width that we will comment next.\\

\begin{figure}[H]
    \centering
    \includegraphics[width=\columnwidth]{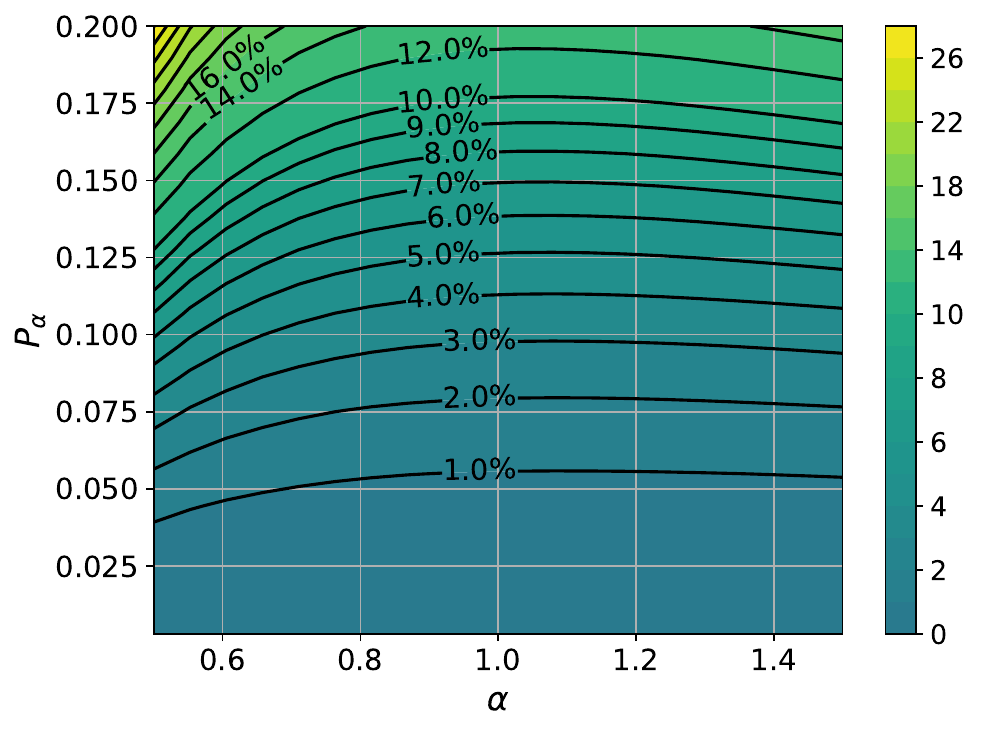}
    \caption{Values of $\epsilon_{\delta_P}$ for the case $a>0$ as a function of $\alpha$ and $P_{\alpha}$. The photon sphere width increase up to $10-20 \%$ with respect to the GR value for values $P_{\alpha} = 0.175 - 0.2$.}
    \label{delta_exp_sol}
\end{figure}

Finally, in Figure \ref{delta_exp_sol}, we plot the deviation with respect to GR of the photon sphere width $\epsilon_{\delta_P}$. As we can see, the correction is always positive in that range and it is of order $1 - 20 \%$ when we vary $P_{\alpha}$ from $0.05 - 0.2$. The larger width of the photon sphere compared to GR arises because, as we have seen, increasing $P_{\alpha}$ causes the photon-sphere radius to decrease faster than the capture radius. As a result, the total reflection trajectories lie further from the photon-sphere radius, where the gravitational field is weaker. This weaker field makes the trajectories more widely separated, leading to a photon-sphere width that is greater than in the standard GR case.

\subsection{Observables for $a<0$}

\begin{figure}
    \centering
    \includegraphics[width=\columnwidth]{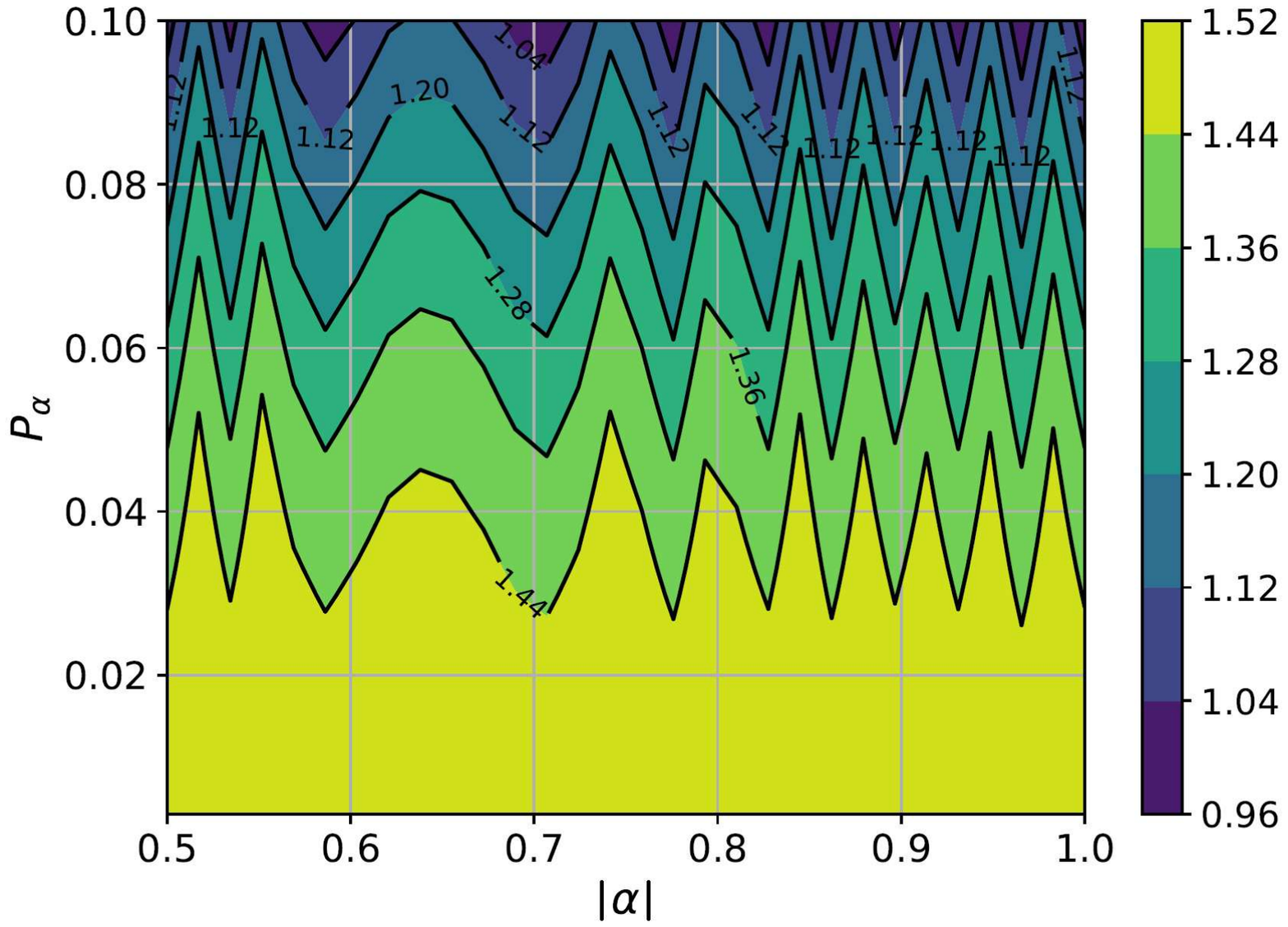}
    \caption{Values of the photon sphere radius $r_P$ for the case $a<0$ as a function of $|\alpha|$ and $P_{\alpha}$. In this situation, the dependence on $\alpha$ has oscillatory behaviour due to $a<0$ but in average $r_P$ decrease with $P_{\alpha}$.}
   
    \label{rp_osc_sol}
\end{figure}

\begin{figure}
    \centering
    \includegraphics[width=\columnwidth]{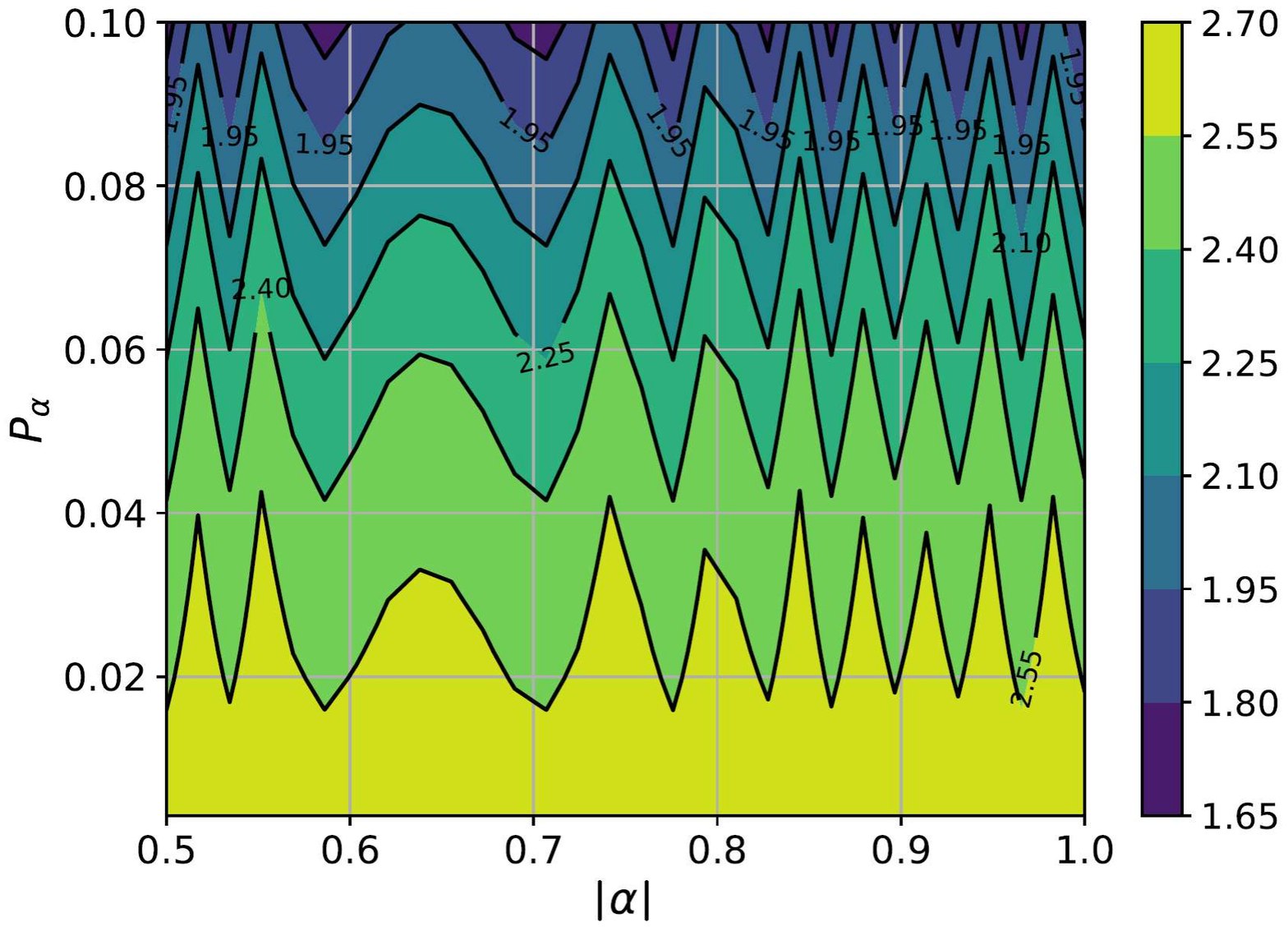}
    \caption{Values of the capture parameter $b_c$ for the case $a<0$ as a function of $|\alpha|$ and $P_{\alpha}$. In this situation, the dependence on $|\alpha|$ has oscillatory behaviour due to $a<0$ but in average $b_c$ decrease with $P_{\alpha}$.}
   
    \label{bc_osc_sol}
\end{figure}

In this case, the dependence on $|\alpha|$ is more subtle, as it modifies the oscillation frequency of the metric. However, for values of $|\alpha|$ of order $1$, the frequency is sufficiently high and the average effect is quite smooth. In Figure \ref{rp_osc_sol} and \ref{bc_osc_sol} we plot the photon sphere radius and the capture parameter respectively. We can see that, even though the values oscillate with $|\alpha|$, average behavior is similar to that of the previous case. As we increase $P_{\alpha}$ the photon sphere radius and the capture parameter decrease. However, as a difference with the previous case, the capture parameter decreases faster. For a $f(R)$ correction up to $10 \%$, the capture parameter decrease from $2.598$ to $1.65$.\\

\begin{figure}
    \centering
    \includegraphics[width=\columnwidth]{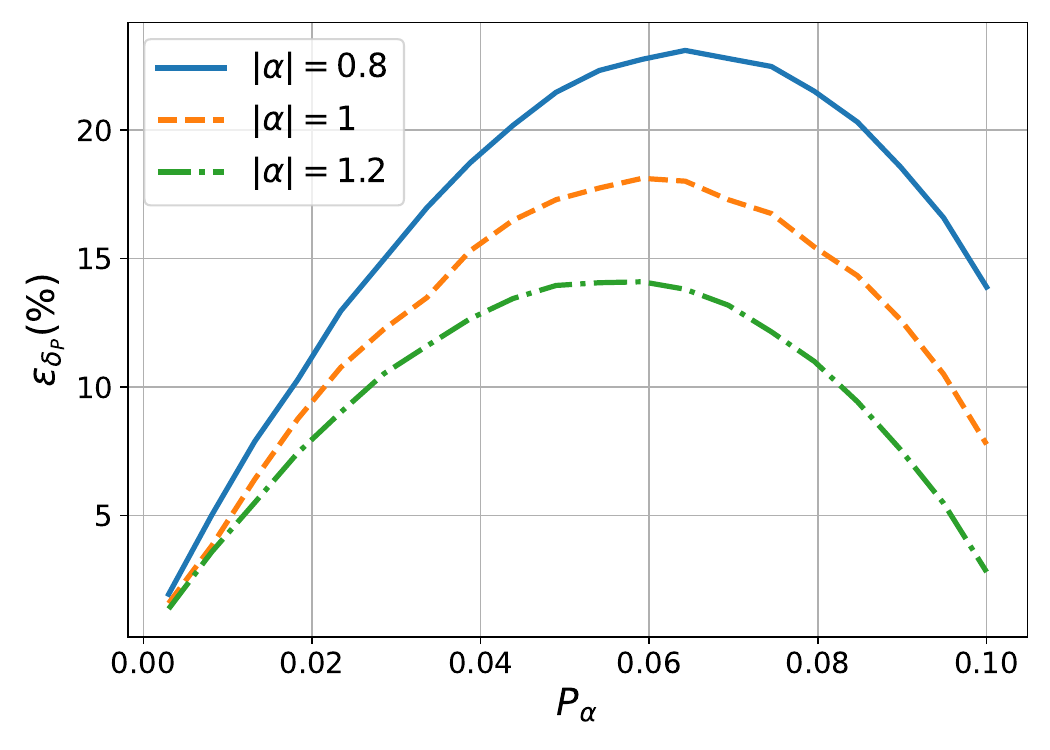}
    \caption{Values of $\epsilon_{\delta_P}$ in the case $a<0$ for values of $|\alpha| = 0.8, 1.0, 1.2$. As we increase $P_{\alpha}$ the photon sphere width increases up to a maximum and then decrease. This is due to the fact that, at some point, the capture parameter decreases faster and approaches to the photon sphere radius.}
   
    \label{delta_osc_sol}
\end{figure}

Finally, in Figure \ref{delta_osc_sol}, we plot the deviation with respect to GR of the photon sphere width $\epsilon_{\delta_P}$ as a function of $P_{\alpha}$ for three different values of $|\alpha|$ near one. In this situation we see that the width increases up to a maximum value and then decrease. This effect is due to that fact that in this case the capture parameter decrease faster. Therefore, the total reflection rings come together, making the width of the photon sphere smaller.\\

\section{Conclusions} \label{conclusions}

In this investigation we have developed a methodology to obtain vacuum solutions in $f(R)$ that are static, spherically symmetric, and that asymptotically recover the Schwarzschild solution of General Relativity. As we are interested in a small correction from GR which might be relevant at high curvatures, we consider a $f(R) = R + a \, R^2$ model and we analyze the differential equation system by expanding around the SW solution at large radii.\\

In this situation, as shown in other studies \cite{Fernandez:2025cfp, AparicioResco:2016xcm}, there are two different solutions depending on whether $a>0$ or $a<0$: an exponential solution or an oscillatory one respectively. We obtain the asymptotic solutions analytically and use them to integrate the differential system numerically from $r\to\infty$ down to $r=0$.\\

These solutions depend on two parameters: in addition to the $f(R)$ parameter $a$, we introduce a parameter that describes the perturbation of the Ricci scalar away from zero. This is due to the fact that, if the Ricci scalar is zero, the SW solution is an exact solution for any value of $a$ in $f(R)$ \cite{delaCruz-Dombriz:2009pzc, delaCruz-Dombriz:2006koo, Casado:2022}. With an appropriate change of variables to work with dimensionless values, the $f(R)$ metric only depends on $\alpha$ and $P_{\alpha}$ parameters. First one is related to $a$ value and describes the scale in with the $f(R)$ correction is dominant in units of the SW radius in GR. Second one is related to the Ricci scalar perturbation from zero and describes the dimensionless correction with respect to GR, this can be seen as an effective dark energy fluid related with the modified gravity theory.\\

The exponential case ($a>0$) for $P(x) > 0$ yields smooth solutions for which $A(x)$ and $B(x)$ go to zero at $x \to 0$, in addition $P(x) \to \infty$. In this work, we do not address the causality properties of the solution; however, this may imply that $r=0$ corresponds to a naked singularity. On the other hand, if we explore $P(x) < 0$ we obtain pathological solutions because all functions diverge as soon as $P(x) \to -2$.\\

The oscillatory case ($a<0$) is more subtle to analyze because the Ricci scalar decays as $1/x$ and not exponentially as in the previous case. This implies that is harder to study numerically in the asymptotic limit. In this case Ricci scalar has damped oscillations that grow as $x \to 0$. $B(x)$ metric function reach $0$ as $A(x)$ goes to $\infty$ so we obtain an effective SW radius. This effective radius is smaller than one, which implies that the photon sphere radius lies closer to the singularity than in GR. Given different values of $\alpha$, this effective SW radius is order $0.7$ for $P_{\alpha} = 0.1$.\\

Once the perturbed solutions to SW in $f(R)$ gravity are obtained, the aim of this work is to analyze strong lensing effects in the  $f(R)$ framework. To this end, we study the photon sphere radius, where light follows an unstable circular orbit, and the capture parameter, i.e., the impact parameter of a light ray that reaches the unstable orbit. Due to the instability of this orbit, this impact parameter also corresponds to the one for which light performs $N \to \infty$ loops before eventually escaping or being absorbed. As has been studied in previous works \cite{Sneppen:2021taq}, when the impact parameter of light approaches the capture parameter $b_c$, there exist certain impact parameters, equally spaced on a logarithmic scale, for which light is either completely reflected or fully transmitted. We revisit this result within the GR framework and extend the analysis to the $f(R)$ case. Moreover, we define the photon-sphere width as $\delta_{P} = b_{0} - b_{c}$, with $b_{0}$ denoting the impact parameter for which light is completely reflected without completing any loops. This work provides a first approximation to the real photon sphere width in the $f(R)$ context. In a practical scenario, simulations accounting for multiple light sources, surrounding dust, and accretion flows are required \cite{Cardenas-Avendano:2022csp, Jia:2024, PhotonRing:2024}. However, this value gives us a reference order of magnitude of the needed sensitivity of experiments like the Event Horizon Telescope, as discussed before, resolving this angular scale requires a precision of order $1 \, \mathrm{\mu as}$.\\

Then we analyze the effect of $f(R)$ perturbed solutions in these parameters. Considering values of $\alpha$ order $1$, which imply that $f(R)$ corrections are dominant at radii $x \approx 1$, and values of $P_{\alpha}$ order $0.05$, we obtain that photon sphere radius and capture parameter decrease with respect to GR, and the photon sphere width increases in a factor $1-10 \, \%$ with respect to the GR case. This occurs in both exponential and oscillating cases. In the context of previous studies that assume Schwarzschild–de-Sitter metrics with constant Ricci scalar \cite{Addazi2021, Nojiri2024, Yue2025, Naskar2025, Jafarzade2024}, our work considers metrics that asymptotically recover the Schwarzschild solution; this requires a model parameter that quantifies deviations of the Ricci scalar from zero, corresponding to an effective dark-energy fluid. Moreover, we define and analyze the \emph{photon sphere width}: the radial distance between the impact parameter of the first total reflection and that of infinitely many reflections, a parameter that would be related with future observables.\\

Finally, we discuss possible directions for future work and potential improvements that could be made in this line of research:

\begin{itemize}
  \item Recent simulations show that the inferred photon-ring radius depends sensitively on plasma properties: at low frequencies it is set by the electron temperature, whereas at higher frequencies the magnetic field is more influential \cite{Desire2025}. Ray-tracing in complex environments also reveals that ultralight boson clouds and plasma variations can cause periodic distortions of the photon ring’s shape and size \cite{Li2025}. Moreover, ray-traced images of thin equatorial disks indicate a degeneracy between spacetime curvature and emission physics, with the peak position of the first photon ring being the most robust observable \cite{Urso2025}.

  \item Our current work assumes static spherically symmetric metrics; an important generalisation is to consider rotating (Kerr-like) solutions in $f(R)$ gravity. Analytical rotating solutions exhibit two horizons and strong central singularities and reduce to the static case when the rotation parameter vanishes \cite{Nashed2021}. Studying how the spin influences the photon-sphere radius, capture parameter and width within our asymptotically Schwarzschild framework will make it possible to compare with observations of rotating black holes.

  \item Finally, we note that the truncated model $f(R) \simeq R + aR^2$ represents only the leading-order term of an analytic expansion around $R = 0$. The pathologies appearing at small radii, such as the divergence of the Ricci scalar or the absence of a regular horizon, highlight the need to include higher-order terms in the curvature. Future work will therefore focus on extending the analysis to models of the form $f(R) = R + aR^2 + bR^3 + \mathcal{O}(R^4)$, in order to investigate whether such corrections can regularize the solutions and restore physically admissible black-hole metrics with proper horizons and singularities. For this purpose, modern horizon-finding algorithms \cite{Etienne2025, Lin2007} could be employed to accurately identify event horizons and curvature singularities in the numerical solutions. Such extensions would also allow for testing whether specific analytic forms of $f(R)$ can reconcile the existence of an event horizon with the asymptotic Schwarzschild behavior.
  
\end{itemize}

To summarize, in this work we have constructed perturbed Schwarzschild solutions in $f(R)$ gravity that asymptotically recover General Relativity, and we have investigated their implications for strong gravitational lensing. In particular, we have analyzed key features of the photon sphere: its radius, the critical capture parameter, and a new quantity that we defined as the photon sphere width. Our results show that even small deviations from General Relativity can produce measurable modifications of these observables, suggesting that precise constraints on $f(R)$-type models may be obtained from high-resolution black hole imaging.

\bibliography{References}
	
\end{document}